\documentclass[final,english]{bullsrsl}
 \pdfoutput=1

\usepackage[T1]{fontenc}

\usepackage{natbib}
\usepackage{graphicx}
\usepackage{xcolor}
\usepackage{comment}
\usepackage{amsmath} 
\usepackage{gensymb} 
\usepackage{subfigure}

\DeclareSymbolFont{matha}{OML}{txmi}{m}{it}
\DeclareMathSymbol{\varv}{\mathord}{matha}{118}

\begin{document}
\title{Runaway O and Be stars found using {\it Gaia} DR3, new stellar bow shocks and search for binaries}

\author[affil={1}]{Mar}{Carretero-Castrillo}
\author[affil={1,2}, corresponding]{Marc}{Ribó}
\author[affil={1}]{Josep M.}{Paredes}
\author[affil={3}]{Paula}{Benaglia}
\affiliation[1]{Departament de Física Quàntica i Astrofísica, Institut de Ciències del Cosmos (ICCUB), Universitat de Barcelona (IEEC-UB)}
\affiliation[2]{Serra H\'unter Fellow}
\affiliation[3]{Instituto Argentino de Radioastronomía (CONICET--CICPBA--UNLP)}

\correspondance{mribo@fqa.ub.edu}
\date{1st October 2024}
\maketitle

\begin{abstract}
A relevant fraction of massive stars are runaways, moving with a significant peculiar velocity with respect to their environment. Kicks from supernova explosions or the dynamical ejection of stars from clusters can account for the runaway genesis. We have used {\it Gaia} DR3 data to study the velocity distribution of massive O and Be stars from the GOSC and BeSS catalogs and identify runaway stars using a 2D-velocity method. We have discovered 42 new runaways from GOSC and 47 from BeSS, among a total of 106 and 69 runaways found within these catalogs, respectively. These numbers imply a percentage of runaways of $\sim$25\% for O-type stars $\sim$5\% for Be-type stars. The higher percentages and higher velocities found for O-type compared to Be-type runaways suggest that the dynamical ejection scenario is more likely than the supernova explosion scenario. We have also performed multi-wavelength studies of our runaways. We have used WISE infrared images to discover 13 new stellar bow shocks around the runaway stars. We have also conducted VLA radio observations of some of these bow shocks. Finally, our runaway stars include six X-ray binaries and one gamma-ray binary, implying that new such systems could be found by conducting detailed multi-wavelength studies. In this work we report on this ongoing project to find new runaway stars, study their interaction with the ISM and search for high-energy binary systems.

\end{abstract}


\section{Introduction} \label{Sect:Introduction}

Massive early-type OB stars are the brightest stars in the Milky Way and play a crucial role in studying the dynamics of young stellar populations, metallicity enrichment, and feedback processes in the interstellar medium (ISM) due to gamma-ray bursts and supernovae \citep{Vanbeveren1998,Woosley2006,Marchant2024}. O-type stars have short lifespans (<10 million years) and powerful stellar winds, while B-type stars can live up to 100 million years. Be stars, a subclass of B stars, exhibit Balmer emission lines and infrared excess due to circumstellar decretion disks \citep{Slettebak1988,Rivinius2013}.

\begin{figure}
\centering
\includegraphics[width=0.6\hsize]{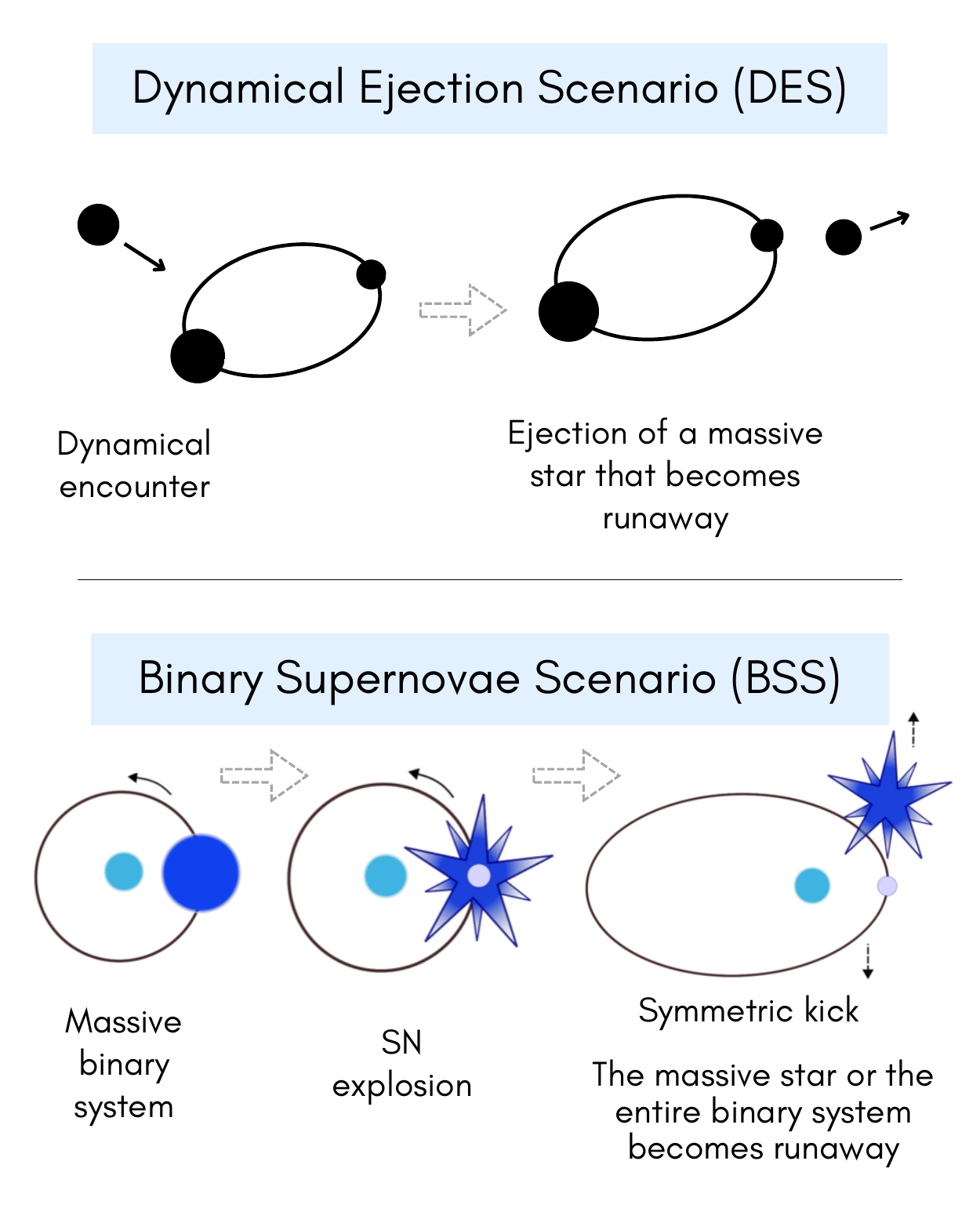}
\bigskip
\begin{minipage}{15cm}
\caption{Different scenarios for the production of massive runaway stars.}
\label{Fig:scenarios}
\end{minipage}
\end{figure}

OB stars are predominantly found in binary systems, with at least 70\% forming interacting binaries \citep{Chini2012,Sana2012}. These systems can evolve into high-mass X-ray binaries (HMXBs), gamma-ray binaries (see \citealt{Dubus2013}), and other exotic systems, enabling the study of non-thermal processes. A significant fraction of O and B stars are classified as runaway stars, exhibiting high peculiar velocities \citep{Blaauw1961,Stone1979,Boubert2018}. Their formation is explained by two scenarios: the dynamical ejection scenario (DES; \citealt{Poveda1967}) and the binary supernova scenario (BSS; \citealt{Blaauw1961}). A two-step ejection process combining both scenarios might also apply in some cases \citep{PflammAltenburg2010}. We show in Fig.~\ref{Fig:scenarios} a schematic representation of the DES and BSS scenarios. We emphasize here that the BSS, in particular, can give rise to the formation of runaway binary systems harboring compact objects, which could eventually be HMXBs or gamma-ray binaries. It is worth noting that some works favor the dominance of the DES scenario \citep{DorigoJones2020} while others favor the dominance of the BSS one \citep{Sana2022}.

\begin{figure}
\centering
\includegraphics[width=0.8\hsize]{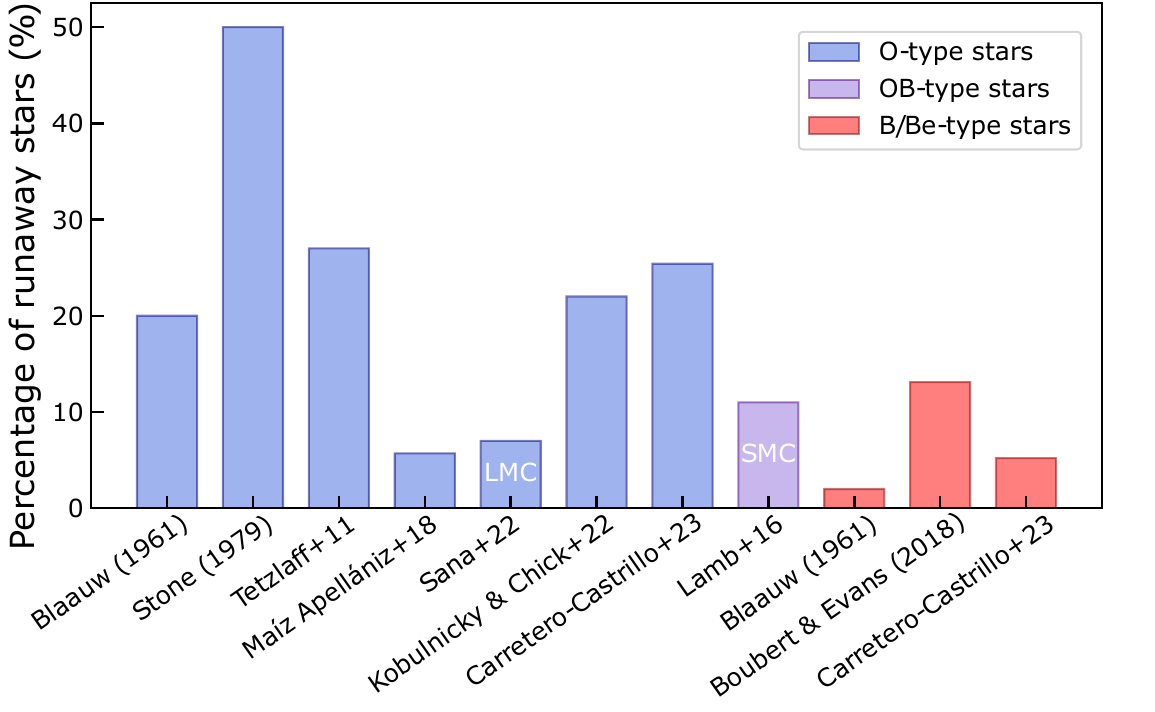}
\bigskip
\begin{minipage}{15cm}
\caption{Histogram of the percentage of runaway stars in the Milky Way, and in the LMC and SMC where indicated.}
\label{Fig:Review}
\end{minipage}
\end{figure}

Extensive research has been conducted to identify massive runaway stars. This includes the use of different data sets: O stars, OB stars, Be stars, young stars with different spectral types, objects in the Milky Way, in the Large Magellanic Cloud (LMC) or in the Small Magellanic Cloud (SMC). In addition, different methods have been used: radial velocities alone, 2D proper motions, 2D velocities, 3D velocities, using thresholds or not, etc. This has an influence in the obtained results. We show the percentage of runaway stars in the Milky Way, LMC and SMC published in several previous works in Fig.~\ref{Fig:Review} and provide more information in Table~\ref{Tab:Review}. As it can be seen, apart from very early studies, the percentage of O-type runaway stars is around 20--30\%, while that of B and Be stars is around 5--10\%. Given the somewhat large dispersions and the publication of {\it Gaia} data with excellent proper motions and parallaxes, it looks natural to conduct a more refined and comprehensive study.

\begin{table}
\centering
\begin{minipage}{15cm}
\caption{Main characteristics of works on all-sky searches of runaways among O, B, Be and young stars in the Milky Way, except for two references of the LMC and SMC.}
\label{Tab:Review}
\end{minipage}
\bigskip

\begin{tabular}{l@{~~~~}c@{~~~~}c@{~~~~}c@{~~~~}c}
\hline
\textbf{Sample}  & 
\textbf{Method} &
\textbf{Threshold}  &
\textbf{Runaways}  &
\textbf{References} \\
\hline
O stars    &  3D vel.  &  30~km~s$^{-1}$  & $\sim$20\% &  \cite{Blaauw1961} \\
O stars    &  3D vel.  &  40~km~s$^{-1}$  & $\sim$49\% &  \cite{Stone1979} \\
$\leq$50~Myr \& \textit{Hipparcos}  &  3D vel.  &  28~km~s$^{-1}$  & $\sim$27\% &  \cite{Tetzlaff2011} \\
O stars \& \textit{Gaia} DR1  &  2D p.m.    &  None  & $\sim$6\% &  \cite{MA2018} \\
O stars in 30 Dor (LMC)             &  1D vel.  &  None  & $\sim$7\% &  \cite{Sana2022} \\
GOSC \& \textit{Gaia} EDR3 &  2D vel.    &  25~km~s$^{-1}$  & $\sim$22\% &  \cite{Kobulnicky2022} \\
GOSC-\textit{Gaia} DR3  &  2D vel.    &  None  & 25.4\% &  \cite{MCC2023} \\
\hline
OB stars in SMC             &  1D vel.    & 30~km~s$^{-1}$  & 11\% &  \cite{Lamb2016} \\

\hline
B stars    &  3D vel.    & 30~km~s$^{-1}$  & $\sim$2\% &  \cite{Blaauw1961} \\
Be \& \textit{Gaia}~DR1 &  3D vel.  &  None  & $\sim$13\% &  \cite{Boubert2018} \\
BeSS-\textit{Gaia} DR3  &  2D vel.  &  None  & ~5.2\% &  \cite{MCC2023} \\
\hline
\end{tabular}
\end{table}

On the other hand, a stellar bow shock forms when a runaway star with a strong stellar wind travels through the ISM, causing ISM matter to accumulate. If the star's velocity exceeds the local sound speed (typically several kilometers per second), supersonic shocks occur around a boundary that separates the stellar wind from the surrounding medium. These structures, consisting of warm dust heated by ultraviolet radiation from the star, emit strongly at infrared wavelengths (IR) and often appear as arc-shaped or bubble-like features, with the leading edge being more pronounced in the direction of the star's motion. They are also potential sites for particle acceleration, where particles may reach relativistic speeds and emit radiation across the electromagnetic spectrum. This possibility has been explored in previous studies, such as in \cite{Benaglia2010}, who detected the first non-thermal radio bow shock. Recent advancements, particularly the improvement of sensitivity in existing radio interferometers in the northern hemisphere and the addition of very sensitive ones in the southern hemisphere, have led to multiple radio detections of bow shocks and have allowed for spectral index studies \citep{VandenEijnden2022a,VandenEijnden2022b,Moutzouri2022}.

In this work we review in Sect.~\ref{Sect:Runaways} the discovery of new massive O and Be runaway stars with \textit{Gaia}~DR3 already published in \cite{MCC2023}, some of which may represent high-mass X-ray or gamma-ray binaries \citep{Ayan2019,Carretero-Castrillo2023}. In addition, we also searched for stellar bow shocks around these runaways (Carretero-Castrillo+24, submitted to A\&A), and present a summary of this work in Sect.~\ref{Sect:BowShocks}. We comment on future prospects to find new binary systems and high-energy sources within our samples in Sect.~\ref{Sect:Binaries}. Finally, we state our conclusions in Sect.~\ref{Sect:Conclusions}.

\section{Search for runaway stars with \textit{Gaia}~DR3}  \label{Sect:Runaways}

\subsection{Data}

{\it Gaia}, a mission by the European Space Agency (ESA) launched in December 2013, is mapping the three-dimensional spatial and velocity distribution of over a billion stars and determining their astrophysical properties \citep{GaiaMission}. Throughout the mission, several {\it Gaia} data releases have progressively improved in astrometry and provided increasing amounts of information (positions, proper motions, parallaxes, temperatures, radial velocities, etc.). For this study, we utilize data from the DR3 release \citep{GaiaDR3}. Specifically, we focus on the stars belonging to two catalogs of massive stars: the Galactic O-Star Catalog (GOSC) \citep{GOSC}, which compiles O-type stars in the Milky Way, and the Be Star Spectra (BeSS) catalog \citep{BeSS}, containing classical Be stars, Herbig Ae/Be stars, and B[e] supergiants. These catalogs are relevant to search for stellar companions of gamma-ray binaries, which are predominantly O and Be-type stars.

We use a version of GOSC containing 643 O and BA stars, but remove two A0 stars and 37 stars flagged as multiple star systems to avoid astrometric problems with {\it Gaia} data, ending with 604 stars. We cross-match their positions with {\it Gaia}-DR3 ones allowing for a maximum separation of 0.5$^{\prime\prime}$, yielding 600 cross matches. However, some of these matches are due to two different {\it Gaia} sources coincident with a single GOSC star. We remove these cases and loose 6 stars, ending up with 598 stars. Afterwards we apply different quality cuts to the resulting catalog to ensure a good quality of the astrometric data and remove: stars in pairs with the same {\it Gaia} \texttt{source\_id}, stars with non 5- or 6-parameter solution, stars with magnitude $G<6$, stars with \texttt{visibility\_periods\_used} $<$ 10, stars with a parallax error over parallax > 0.2, stars with negative parallaxes, and stars with RUWE parameter > 1.35. We also reject 6 stars with galactocentric distances smaller than 5~kpc because of the validity range of the Galactic rotation curve used. This yields a subset of 417 stars stars, which we call the GOSC-\textit{Gaia}~DR3 catalog.
For the BeSS catalog, containing 2330 Be stars, we conduct a similar process and remove Herbig Ae/Be stars and B[e] supergiants, stars in the LMC or SMC, we use a maximum separation for the cross match of 1.5$^{\prime\prime}$ due to the limited astrometric quality in the BeSS database, and we apply the same quality cuts mentioned above. We end up with a refined catalog of 1335 stars that we call the BeSS-\textit{Gaia}~DR3 catalog.

\begin{figure}
\centering
\includegraphics[width=0.75\hsize]{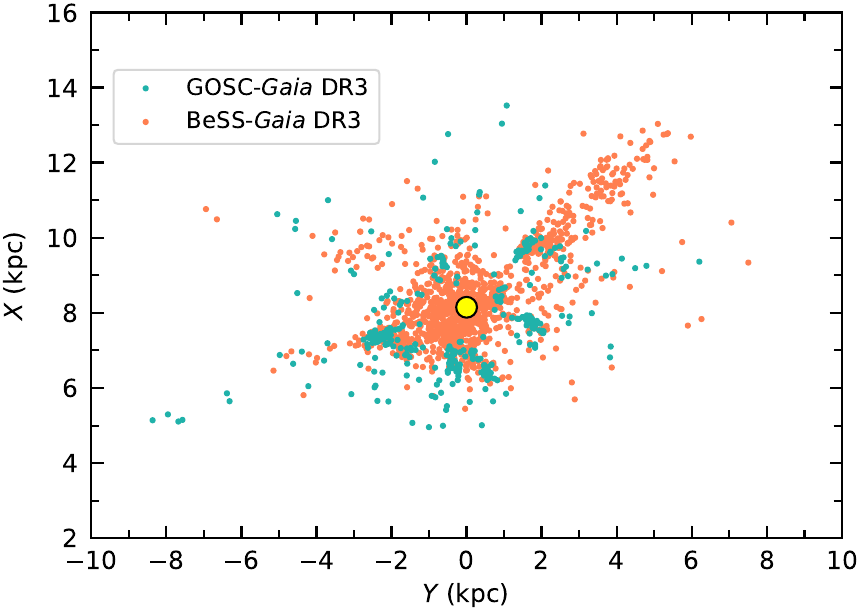}
\bigskip
\begin{minipage}{15cm}
\caption{Galactocentric $(X,Y)$ distributions of the GOSC-\textit{Gaia}~DR3 stars (light green), and the BeSS-\textit{Gaia}~DR3 stars (orange). The position of the Sun is marked with a yellow circle at $\left(X, Y\right) = \left(8.15,0\right)$~kpc. Adapted from \cite{MCC2023}.}
\label{Fig:XYGal}
\end{minipage}
\end{figure}

{\it Gaia} DR3 provides very accurate positions, proper motions and parallaxes of the stars under study. To derive their distances and related uncertainties we use the procedure described in \cite{MA2022}. The Galactocentric $(X,Y)$ distributions of the stars in the GOSC- and BeSS-\textit{Gaia}~DR3 catalogs are presented in Fig.~\ref{Fig:XYGal}. We note that the 
overdensity of BeSS-\textit{Gaia}~DR3 stars in the direction from the Sun toward $(X,Y) = (13,5)$~kpc is due to low-resolution surveys of Be stars in that particular direction.

\subsection{Method}

\begin{figure}[t!]
\centering
\includegraphics[width=0.5\hsize]{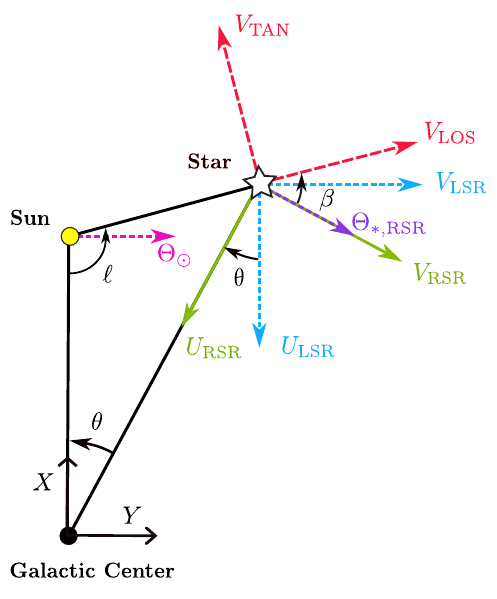}
\bigskip
\begin{minipage}{15cm}
\caption{Reference systems used in this work, depicted in the Galactic $XY$ plane. The Sun and the Galactic Center (GC) are indicated with a yellow and black circle, respectively. The white star represents a given star in our catalogs. $l$ is the Galactic longitude, $\theta$ is the angle between the star and the Sun as seen from the GC and $\beta$ is a rotation angle that we use to shift from RSR to the new reference system. The Galactic circular rotation velocity is $\Theta_\odot$ at the position of the Sun and $\Theta_{*,{\rm RSR}}$ at the position of the star. LSR and RSR velocities are shown in light blue and green, respectively. The tangential velocity $V_\text{TAN}$ and the line-of-sight velocity $V_\text{LOS}$ are shown in red. The third velocity components of the LSR and the RSR systems, $W_\text{LSR}$ and $W_\text{RSR}$, are not shown in the figure because they are perpendicular to this plane. Adapted from \cite{MCC2023}.}
\label{Fig:ReferenceSystems}
\end{minipage}
\end{figure}

To detect runaway stars, we first need to compute their velocities. To this end we need positions, parallaxes, proper motions and radial velocities. However,  {\it Gaia}~DR3 does not provide many radial velocities for O and Be stars. Therefore, we assume a Galactic rotation curve (model A5 of \citealt{Reid2019}) from which we compute a theoretical radial velocity (i.e., the radial velocity the star would have if it was moving following the Galactic rotation curve). We then apply the usual reference system transformations from the Local Standard of Rest (LSR) to the Regional Standard of Rest (RSR) from \cite{Johnson&Soderblom}. Finally, we define a new reference system, composed of the tangential velocity $V_{\text{TAN}}$, which is contained in the plane of the sky and is parallel to the Galactic plane, the line of sight velocity $V_{\text{LOS}}$, and the velocity component $W_{\text{RSR}}$ that is perpendicular to this plane. The different reference systems are shown in Fig.~\ref{Fig:ReferenceSystems}. We finally use the 2D velocity system composed of $V_{\text{TAN}}$ and $W_{\text{RSR}}$. For further details about the methodology and related propagation of uncertainties we refer the reader to \cite{MCC2023}.

Second, we need to define a runaway criterion. We classified the runaway stars based on how significant the 2D velocities of the stars are with respect to the mean Galactic rotation.
To do this, we used the $E$ parameter presented in Eq.~\ref{Eq:Criteria}, which takes into account the means and standard deviations of Gaussian fits to the velocity distributions of the field stars ($\mu_{V_{\text{TAN}}}^{\text{GF}}, \mu_{W_{\text{RSR}}}^{\text{GF}}, \sigma_{V_\text{TAN}}^{\text{GF}}, \sigma_{W_\text{RSR}}^{\text{GF}}$), the individual errors of the stars ($\sigma_{V_{\text{TAN},*}}, \sigma_{W_{\text{RSR},*}}$), and is normalized to a 3-sigma-confidence level. Therefore, after applying an iterative 3-$\sigma$ clipping process, stars with $E>1$ are classified as runaway stars, while stars with $E<1$ are classified as field stars.

\begin{equation}
    \label{Eq:Criteria}
        E =\sqrt{ \left(\frac{V_{\text{TAN}} - \mu_{V_{\text{TAN}}}^{\text{GF}}}{3\sqrt{{\sigma_{V_\text{TAN}}^{\text{GF}}}^2 + \sigma_{V_{\text{TAN},*}}^2}}\right)^2 + \left(\frac{W_{\text{RSR}} - \mu_{W_{\text{RSR}}}^{\text{GF}}}{3\sqrt{{\sigma_{W_\text{RSR}}^{\text{GF}}}^2 + \sigma_{W_{\text{RSR},*}}^2}}\right)^2
        }.
\end{equation}

\subsection{Results and discussion}

The means and standard deviations obtained from the final Gaussian fits to the velocity distributions of the field stars, which are used in Eq.~\ref{Eq:Criteria}, are presented in Table~\ref{Tab:FitsAfterClippingField}. Notably, $\sigma_{V_{\text{TAN}}}^{\text{GF}}$ is greater than $\sigma_{W_{\text{RSR}}}^{\text{GF}}$ for both catalogs, with BeSS-\textit{Gaia}~DR3 stars exhibiting a higher $\sigma_{V_{\text{TAN}}}^{\text{GF}}$ than those from GOSC-\textit{Gaia}~DR3 because they are more affected by Galactic velocity diffusion in the disk (see \citealt{MCC2023} for additional checks to reach this conclusion).

\begin{table}
\centering
\begin{minipage}{15cm}
\caption{Means and standard deviations of the velocity distributions for the field stars. Adapted from \cite{MCC2023}.}
\label{Tab:FitsAfterClippingField}
\end{minipage}
\bigskip

\begin{tabular}{l@{~~~~}c@{~~~~}cr@{~~$\pm$~~}lr@{~~$\pm$~~}l}
\hline
\textbf{Catalog}  & \textbf{Stars} & \textbf{Field Stars} & 
$\mathbf{\mu_{V_{\text{TAN}}}^{\text{GF}}}$ & $\mathbf{\sigma_{V_{\text{TAN}}}^{\text{GF}}}$ &
$\mathbf{\mu_{W_{\text{RSR}}}^{\text{GF}}}$ & $\mathbf{\sigma_{W_{\text{RSR}}}^{\text{GF}}}$
\\
 & \# & \#
 & \multicolumn{2}{c}{(km~s$^{-1}$)} 
 & \multicolumn{2}{c}{(km~s$^{-1}$)}
\\
\hline
GOSC-\textit{Gaia}~DR3              & ~~417 & ~~311 &     1.0 & 6.6   &  $-$1.0 & 5.3 \\
BeSS-\textit{Gaia}~DR3              &  1335 &  1266 &     1.6 & 9.3   &  $-$0.5 & 4.9 \\
\hline
\end{tabular}
\end{table}

\begin{table}
\centering
\begin{minipage}{15cm}
\caption{Means and standard deviations of the velocity distributions for the runaway stars. Adapted from \cite{MCC2023}.}
\label{Tab:FitsAfterClippingRun}
\end{minipage}
\bigskip

\begin{tabular}{l@{~~~~}c@{~~~~}cr@{~~$\pm$~~}lr@{~~$\pm$~~}l}
\hline
\textbf{Catalog}  & \textbf{Stars} & \textbf{Runaway Stars} &
$\mathbf{\mu_{V_{\text{TAN}}}}$ & $\mathbf{\sigma_{V_{\text{TAN}}}}$ &
$\mathbf{\mu_{W_{\text{RSR}}}}$ & $\mathbf{\sigma_{W_{\text{RSR}}}}$ 
\\
 & \# & \#
 & \multicolumn{2}{c}{(km~s$^{-1}$)} 
 & \multicolumn{2}{c}{(km~s$^{-1}$)}
\\
\hline
GOSC-\textit{Gaia}~DR3              & ~~417 &  106 &  $-$3.1 & 39.7  &  1.1 & 38.7 \\
BeSS-\textit{Gaia}~DR3              &  1335 & ~~69 &  $-$7.7 & 34.6  &  1.0 & 22.3 \\
\hline
\end{tabular}
\end{table}

\begin{figure}
\centering
\includegraphics[width=0.65\hsize]
{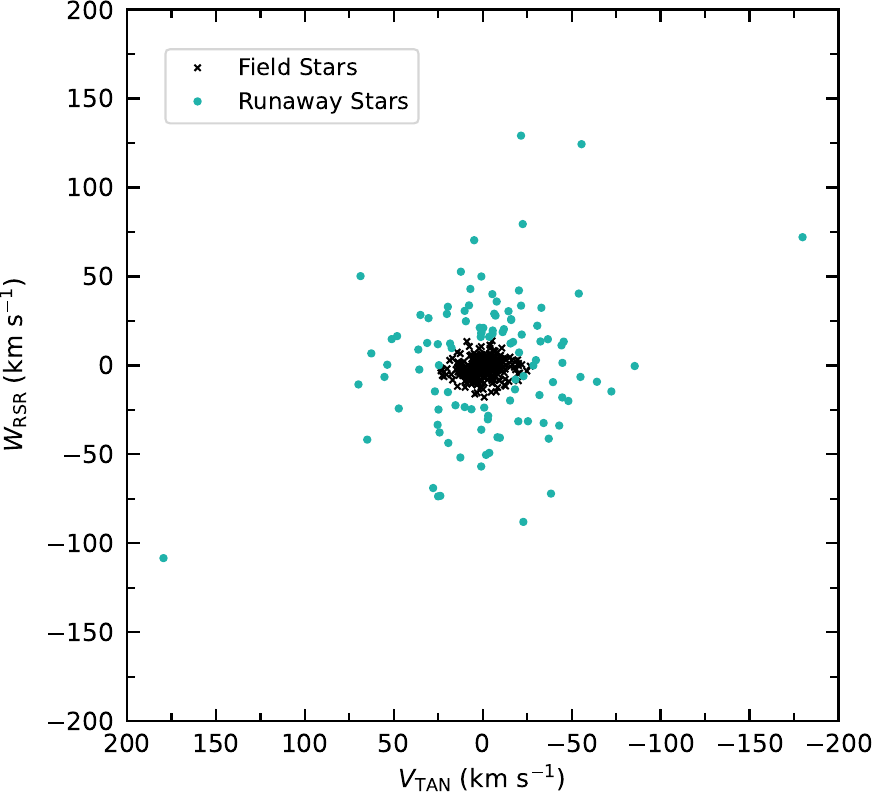}
\bigskip
\begin{minipage}{15cm}
\caption{$W_{\text{RSR}}$ as a function of $V_{\text{TAN}}$ for the GOSC-\textit{Gaia}~DR3 stars. Field stars are depicted in black, and runaway stars are shown in light green. Adapted from \cite{MCC2023}.}
\label{Fig:2D_GOSCGaiaDR3}
\end{minipage}
\end{figure}

\begin{figure}
\centering
\includegraphics[width=0.65\hsize]
{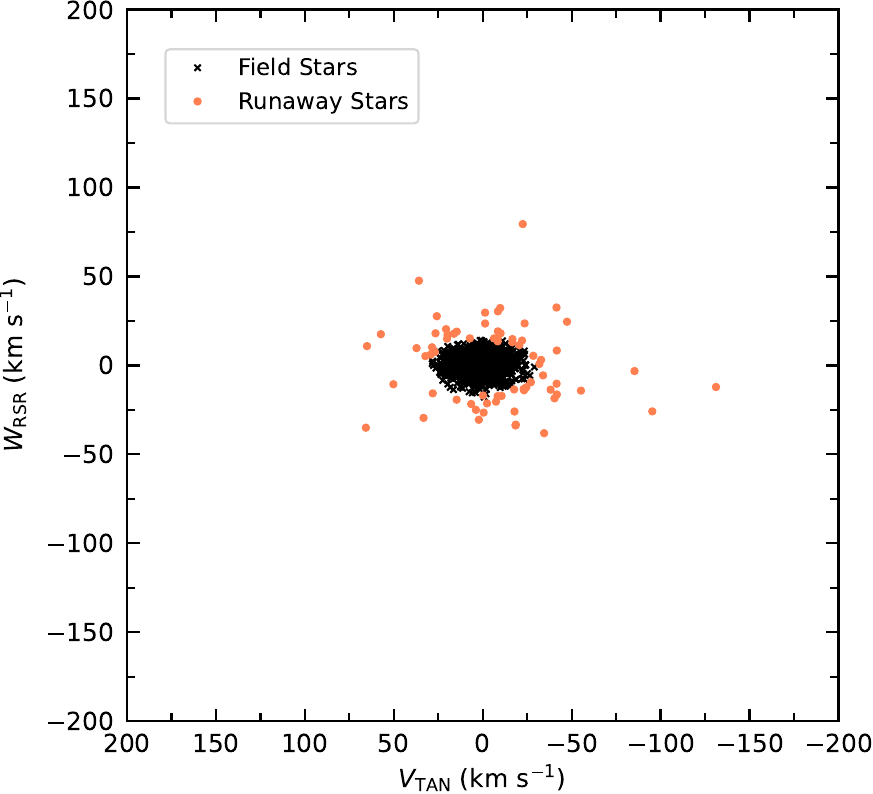}
\bigskip
\begin{minipage}{15cm}
\caption{$W_{\text{RSR}}$ as a function of $V_{\text{TAN}}$ for the BeSS-\textit{Gaia}~DR3 stars. Field stars are depicted in black, and runaway stars are shown in orange. Adapted from \cite{MCC2023}.}
\label{Fig:2D_BeSSGaiaDR3}
\end{minipage}
\end{figure}

After applying the methodology described in the previous subsection, we identified 106 runaway stars in the GOSC-\textit{Gaia}~DR3 catalog (42 newly discovered) and 69 in the BeSS-\textit{Gaia}~DR3 catalog (47 newly discovered), representing 25.4\% and 5.2\% of the catalogs, respectively. The means and standard deviations (without applying Gaussian fits) to the velocity distribution of the runaway stars are presented in Table~\ref{Tab:FitsAfterClippingRun}. Runaway stars exhibit high velocity dispersions of 20--40 km~s~$^{-1}$, consistent with literature values \citep{Stone1991,Tetzlaff2011}. Figures~\ref{Fig:2D_GOSCGaiaDR3}~and~\ref{Fig:2D_BeSSGaiaDR3} show the 2D $\left(V_{\text{TAN}},W_{\text{RSR}}\right)$ velocity distributions for the GOSC- and BeSS-\textit{Gaia}~DR3 catalogs, respectively, including a color distinction between runaways and field stars. Field stars are clustered around (0,0) while runaway stars present higher velocities. In particular, the velocities of the O runaway stars are higher than those of Be runaways. The lower standard deviations in $W_{\text{RSR}}$ together with the larger uncertainties in  $V_{\text{TAN}}$ imply that more runaways are identified in the $W_{\text{RSR}}$ component.

\begin{figure}
\centering
\includegraphics[width=1\hsize]
{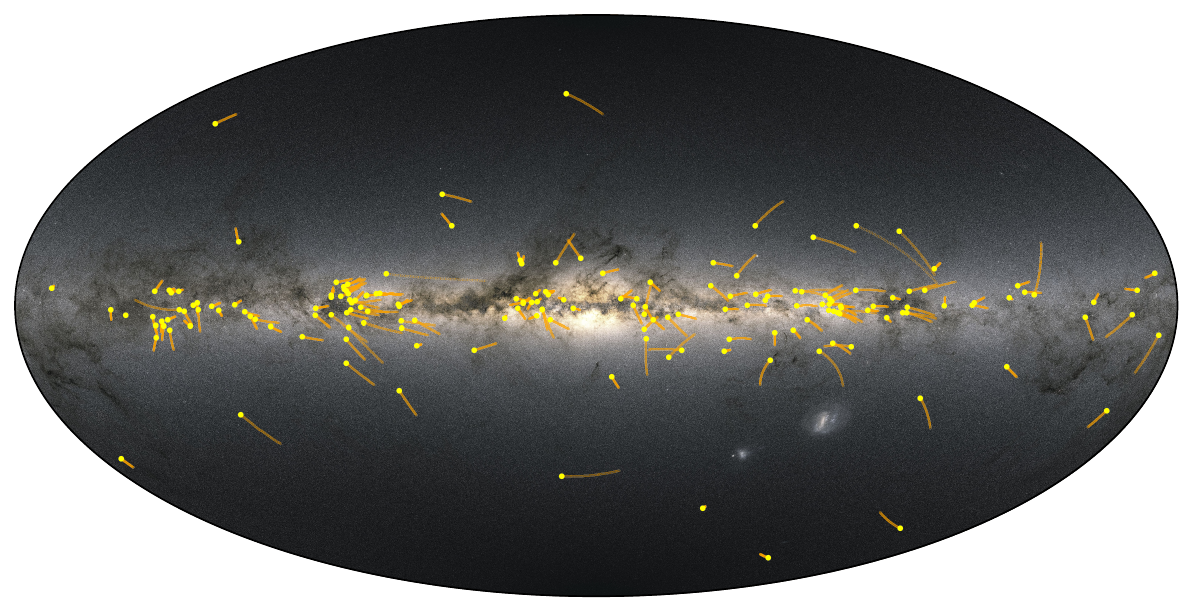}
\bigskip
\begin{minipage}{15cm}
\caption{All-sky view in Galactic coordinates showing the O and Be runaway stars found in the GOSC- and BeSS-\textit{Gaia}~DR3 catalogs as yellow circles. Orange traces indicate the positions of the stars up to 3~Myr in the future. The background sky map of the Milky Way is that of \textit{Gaia}~DR3 (credits: ESA/\textit{Gaia}/DPAC).}
\label{Fig:runawayDistribution}
\end{minipage}
\end{figure}

The positions in Galactic coordinates of both the O and Be runaway stars are shown as yellow circles in Fig.~\ref{Fig:runawayDistribution}, together with their positions up to 3~Myr in the future indicated with orange traces. The runaway stars present 2--3 times more dispersion in $b$ than the field stars (not shown here), which is expected given that they have been expelled from their birthplaces during their formation as runaways.

We also classified the runaways in different sub-spectral type bins and found that the runaway percentage decreases as we move to later spectral types. Table~\ref{Tab:SpectralType} shows the results in number and percentage. The higher percentages and higher velocities found for O-type stars compared to Be-type stars are in favor of the dominance of the DES over the BSS scenario.

\begin{table}[t!]
\centering
\begin{minipage}{15cm}
\caption{Field and runaway stars as a function of spectral type. Adapted from \cite{MCC2023}.}
\label{Tab:SpectralType}
\end{minipage}
\bigskip

\begin{tabular}{lcccc}
\hline
\textbf{Spectral Type}  & 
\textbf{Stars}          &
\multicolumn{2}{c}{\textbf{Runaway Stars}} \vspace{-4mm} \\
\vspace{-2mm}
          &       & \multicolumn{2}{c}{-----------------------} 
\vspace{0mm}\\
          &  \#   & \#   & \%      \\
\hline
O2--O7    &  199  &  50  &   25.1  \\
O8--O9    &  194  &  46  &   23.7  \\
B0e--B3e  &  585  &  36  &  ~~6.2  \\
B4e--B9e  &  482  &  23  &  ~~4.8  \\
\hline
\end{tabular}
\end{table}

\section{Search for stellar bow shocks} \label{Sect:BowShocks}

We conducted a search for stellar bow shocks around the runaways found in \cite{MCC2023} using data from the Wide-field Infrared Survey Explorer (WISE) \citep{WISE2010}, which is a NASA-funded mission designed to survey the entire sky in the mid-infrared. Completed in 2010, WISE provided significantly higher sensitivity than previous infrared surveys, mapping the sky in multiple bands with varying resolutions. 

Bow shocks are associated with warm dust emission, and this emission is primarily expected in the WISE W4 band (22~$\mu$m). Therefore, we utilized the W4 band to search for bow shock or bubble-like structures around runaway stars. In addition, we corrected the {\it Gaia} DR3 proper motions to account for the ISM motion caused by Galactic rotation. For this, we used the prescription provided in \cite{Comeron2007}. In this way, we can know the actual motion of the runaway stars with respect to the bow shock candidate. Therefore, we classified an IR structure as a stellar bow shock or bubble if it was close to the position of the star and showed an arc-shaped structure/or the rim of the bubble in the direction of the runaway star’s corrected proper motion.

\begin{figure}
\centering
\includegraphics[width=0.6\hsize]{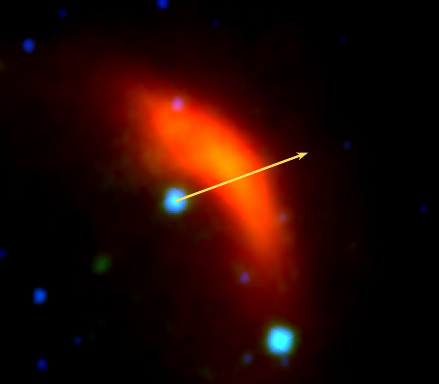}
\bigskip
\begin{minipage}{15cm}
\caption{WISE image of a known bow shock as an example.}
\label{Fig:BS_example}
\end{minipage}
\end{figure}

We found 13 new stellar bow shock or bubble candidates, mostly associated with O-type stars. We also identified 16 known bow shocks or bubble candidates discovered in previous works. Figure~\ref{Fig:BS_example} presents an example of a WISE RGB image in W4+W3+W2 of one of these known bow shocks. Additionally, we introduced a new category of "mini-bubble" sources which are around twice the W4 PSF, with two examples linked to Be-type stars. We performed an IR geometrical characterization of the bow shocks and bubbles, measuring projected distances such as the stand-off distance $R$, the length, and the width. Using these measurements, we estimated the ISM densities at the bow shock positions using the following equation adapted from \cite{Wilkin1996}:
\begin{equation} \label{Eq:nISM}
    R_0 = \sqrt{\frac{\dot{M}\varv_\infty}{4\pi\mu n_{\text{ISM}} {V_\text{PEC}^\text{3D}}^2}}
\end{equation}
where $R_0$ is the stand-off distance, $\dot{M}$ is the mass-loss rate, $\varv_\infty$ is the wind terminal velocity, $\mu$ is the mass of the ISM per H atom, and $V_\text{PEC}^\text{3D}$ is the three dimensional (3D) peculiar velocity of the star.
Using the projected stand-off distance $R$ instead of $R_0$,  we found a median $n_{\text{ISM}}$ of $\sim$4~cm$^{-3}$. This estimate aligns with previous works, though in some particular cases we obtain very-high density values that may be overestimated due to projection effects.

We also searched for radio emission from the bow shocks using data from surveys like NVSS, VLASS, and RACS. While we found radio emission coincident with the bow shock position for some sources, the radio morphology differed from the one of the IR bow shock, making it difficult to confirm radio counterparts. We explored both thermal and non-thermal mechanisms as potential origins of the radio emission and predicted the radio flux density for undetected bow shocks as done in \cite{VandenEijnden2022b}. Despite the non-detection of the sources, they remain consistent with the possibility to display non-thermal radio emission that could be detected in deeper surveys.

Overall, our results highlight the need for more sensitive radio observations to confirm bow shocks at radio wavelengths and investigate the physical processes driving their radio and high-energy emissions. Instruments like the Very Large Array (VLA) could play a crucial role in uncovering a new population of radio bow shocks. As work in progress, we are now analyzing VLA data of eight bow shock candidates found in this work.

\section{Search for binaries} \label{Sect:Binaries}

As mentioned in the introduction, the BSS can result in systems such as HMXBs and gamma-ray binaries. We looked for known of these sources among our runaways and found six HMXBs and one gamma-ray binary. In particular, we identified the gamma-ray binary LS~5039 and the HMXBs Vela~X-1, 4U~1700$-$377, IGR~J08408$-$4503, and Cygnus X$-$1 in the GOSC-\textit{Gaia}~DR3 catalog. From the BeSS-Gaia DR3 runaways, we found the SAX~J2103.5+4545 and V~0332+53 HMXBs. Notably, Vela~X-1 and SAX~J2103.5+4545 are identified here as runaway HMXBs for the first time.

As an outlook, on the one hand, we want to search for binarity hints among our runaways to create a reduced list of runaway binary systems candidates. To this end, we will check for radial velocity variations and signatures of fast rotation \citep{Britavskiy2023,Holgado2022}. On the other hand, since HMXBs and gamma-ray binaries present emission in different wavelengths, we are conducting a multi-wavelength study (from radio to high-energy gamma rays) of this sample to create a reduced list of high-energy binary candidates. Once we have a reduced list of interesting objects, we plan to prepare observations of the most promising candidates to be HMXBs and gamma-ray binaries. This could potentially help to answer open questions for gamma-ray binaries, of which we only know nine systems at the time of writing.

\section{Conclusions} \label{Sect:Conclusions}

In this work we have summarized the status of our ongoing project to find new runaway stars, study their interaction with the ISM and search for high-energy binary systems. In particular, we reach the following conclusions for each topic:

\begin{itemize}
    \item Ruanaway stars. After using {\it Gaia} DR3 data and the GOSC and BeSS catalogs we have detected 175 runaways, 89 of which are new discoveries. We have found that $\sim$25\% of O-type stars from our sample are runaways, while for Be-type stars this decreases to $\sim$5\%. O-type stars runaways have higher velocities than B-type ones. These results suggest that the dynamical ejection scenario is more likely than the supernova explosion scenario.
    \item Bow shocks. We have used WISE infrared images to discover 13 new stellar bow shocks and bubbles around the runaway stars. We have characterized their morphology and derived ISM densities. The analysis of VLA observations of some of these sources is ongoing.
    \item High-energy binary systems. Among the sample of runaways stars we have found one gamma-ray binary and six HMXBs. Ongoing multi-wavelength studies to search for binarity and high-energy emission could provide the discovery of new such systems. 
\end{itemize}

\begin{acknowledgments}

We would like to thank the members of the \textit{Gaia} team at Universitat de Barcelona for many useful discussions. We thank J. Maíz Apellániz, S.R. Berlanas, S. Sim\'on-D\'{\i}az, C. Martínez-Sebastián, M. Pantaleoni González, G. Holgado and A. de Burgos for valuable insights and feedback.
We acknowledge financial support from the State Agency for Research of the Spanish Ministry of Science and Innovation under grants, PID2022-136828NB-C41/AEI/10.13039/\\501100011033/ERDF/EU, and PID2022-138172NB-C43/AEI/10.13039/ 501100011033/ERDF/\\EU, and through the Unit of Excellence María de Maeztu 2020-2023 award to the Institute of Cosmos Sciences (CEX2019-000918-M). We acknowledge financial support from Departament de Recerca i Universitats of Generalitat de Catalunya through grant 2021SGR00679. MC-C acknowledges the grant PRE2020-094140 funded by MCIN/AEI/10.13039/501100011033 and FSE/ESF funds.
Fig.~\ref{Fig:runawayDistribution} was generated using the \texttt{mw-plot} Python package. This work has made use of data from the European Space Agency (ESA) mission {\it Gaia} (\url{https://www.cosmos.esa.int/gaia}), processed by the {\it Gaia} Data Processing and Analysis Consortium (DPAC, \url{https://www.cosmos.esa.int/web/gaia/dpac/consortium}). Funding for the DPAC has been provided by national institutions, in particular the institutions participating in the {\it Gaia} Multilateral Agreement.
This work has made use of the BeSS database, operated at LESIA, Observatoire de Meudon, France: \url{http://basebe.obspm.fr}.
This publication makes use of data products from the Wide-field Infrared Survey Explorer, which is a joint project of the University of California, Los Angeles, and the Jet Propulsion Laboratory/California Institute of Technology, funded by the National Aeronautics and Space Administration.
This research has made use of NASA’s Astrophysics Data System.
This research has made use of the SIMBAD database, operated at CDS, Strasbourg, France.

\end{acknowledgments}

\begin{furtherinformation}

\begin{orcids}
\orcid{0000-0002-1426-1311}{Mar}{Carretero-Castrillo}
\orcid{0000-0001-6944-6383}{Marc}{Rib\'o}
\orcid{0000-0002-1566-9044}{Josep M.}{Paredes}
\orcid{0000-0002-6683-3721}{Paula}{Benaglia}

\end{orcids}

\begin{authorcontributions}
MC-C: data curation, formal analysis, investigation, methodology, project administration, software, visualization, writing-original draft;\\
MR: conceptualization, funding acquisition, investigation, methodology, supervision, writing-original draft;\\
JMP: conceptualization, funding acquisition, investigation, supervision;\\
PB: conceptualization, data curation, formal analysis, investigation.\\
\end{authorcontributions}

\begin{conflictsofinterest}
The authors declare no conflict of interest.
\end{conflictsofinterest}

\end{furtherinformation}

\bibliographystyle{bullsrsl-en}

\bibliography{mybibliography}

\end{document}